\documentclass[iop]{emulateapj}

\usepackage{natbib,aas_macros}
\usepackage{color}
\usepackage{ulem}
\usepackage{dcolumn}
\usepackage{amsmath}
\usepackage[flushleft]{threeparttable}

\renewcommand{\apjl}{ApJL}

 \shorttitle{$\alpha$QSE Nucleosynthesis}
 \shortauthors{S.Fujibayashi, T.Yoshida and Y.Sekiguchi}

\begin{document}

\title{Alpha-constrained QSE nucleosynthesis in high-entropy and fast-expanding material}

\author{
Sho~Fujibayashi
}
\affil{
Department of Physics, Graduate School of Science, Kyoto University, Kyoto 606-8502, Japan
}
\email{
fujibayashi@tap.scphys.kyoto-u.ac.jp
}

\author{
Takashi~Yoshida
}
\affil{
Department of Astronomy, Graduate School of Science, University of Tokyo, Tokyo 113-0033, Japan
}

\author{
Yuichiro~Sekiguchi
}
\affil{
Department of Physics, Toho University, Funabashi, Chiba 274-8510, Japan
}


\begin{abstract}
We investigate the nucleosynthesis process in high-entropy ($s/k_{\rm B}\gtrsim100$) and very fast-expanding ($\tau_{\rm exp}\sim10^{-3}\ {\rm s}$) materials.
In such a material with the electron fraction near 0.5, an interesting nucleosynthesis process occurs.
In this process, the abundance distribution of heavy-nuclei of $A>100$ achieve quasi-statistical equilibrium (QSE) at high temperature and the abundances are frozen at the end of the nucleosynthesis.
We explain this abundance distribution using the ``alpha-constrained QSE" abundances formulated in this paper.
We demonstrate that this nucleosynthesis would occur in neutrino-driven winds from massive proto-neutron stars in hypernovae, where $A\sim140$ $p$-nuclei are synthesized.
\end{abstract}

\section{introduction}
The origins of the proton-rich ($p$-) nuclei have been under debate for a long time.
They are not synthesized by neutron-capture reactions because unstable neutron-rich nuclei are prevented to decay into $p$-nuclei by underlying stable nuclei.
There are some sites proposed for these $p$-nuclei synthesis.
The most successful process that makes these $p$-nuclei would be the $\gamma$-process in CCSNe or these progenitors, in which $p$-nuclei are produced by photo-disintegration of the pre-existing heavy-nuclei.
However, heavy $p$-nuclei of $110<A<125$ are underproduced in this process \citep{1995A&A...298..517R,2002ApJ...576..323R,2003PhR...384....1A}.
The $\nu p$-process in the proton-rich material in the neutrino-driven wind (NDW) in SNe would produce these $p$-nuclei \citep{2006PhRvL..96n2502F,2006ApJ...644.1028P,2006ApJ...647.1323W,2011ApJ...729...46W}.
In the $\nu p$-process, the heavy-nuclei are produced through $(p,\gamma)$ and successive $(n,p)$ reactions (neutrons are supplied via $p+\bar{\nu}_e \rightarrow n+e^+$ reactions).

\cite{2004ApJ...617L.131J} suggested that these light $p$-nuclei could be produced in a fast-expanding, high-entropy, and slightly proton-rich material.
They argued that this is due to the lack of the alpha particle abundance, because the reactions which produce alpha particles from nucleons proceed slower than the expansion of the material \citep[see also][]{2002PhRvL..89w1101M}.
However, a short expansion timescale of $\sim 10^{-4}$ s have to be realized for the occurrence of such a nucleosynthesis, and whether it would occur in some real astronomical phenomena has been unclear.

\cite{2015ApJ...810..115F} suggested that a NDW from temporarily formed massive proto-neutron star (PNS) in a hypernova (HN)  would be a possible site for such a nucleosynthesis.
They showed that the {\it HN wind} has much faster expansion velocity than that of a NDW in an ordinary SN because the massive PNS has very high neutrino luminosity.
As a result, light $p$-nuclei production indeed occurs in the HN wind with realistic parameters.
Therefore, it is worth closely investigating the synthesis process of these heavy-element and $p$-nuclei.

In the materials in the NDW, due to the high entropy and the high expansion velocity, the triple-alpha reaction does not proceed sufficiently so that alpha-particles remain even when the temperature decreases to $\sim 6\times 10^9$ K. (This is so called ``alpha-rich freeze-out.")
\cite{1998ApJ...498..808M} showed that the abundances in such materials are well described by the quasi-statistical equilibrium (QSE) abundances, which minimize the free energy in the system under the constraint that the total number of heavy-nuclei (the nuclei except neutron, proton, and alpha-particle) is fixed.

Taking the study into account, we set our goal in this paper as formulating the nucleosynthesis in high-entropy and fast-expanding materials using the abundance in a special case of QSE where alpha-particles and nucleons are not in equilibrium.
We call the special QSE the {\it alpha-constrained QSE ($\alpha$QSE)}.
In the following, we call the abundance in the $\alpha$QSE the $\alpha$QSE abundance, and the nucleosynthesis via the $\alpha$QSE abundance the $\alpha$QSE nucleosynthesis.
In Section 2, we formulate the $\alpha$QSE abundance and show distributions in some cases.
In Section 3, we compare the $\alpha$QSE abundance and the abundance dynamically calculated with nuclear reaction network, and discuss the condition for the $\alpha$QSE nucleosynthesis.
Then, in Section 4, we show the $\alpha$QSE nucleosynthesis indeed occurs in the HN wind, and discuss the points different from the nucleosynthesis in the NDW in the ordinary SNe.
Finally discussion and conclusion are given in Sections 5 and 6, respectively.

\section{Theory of $\alpha$QSE}
In this section we derive the abundance distribution where the alpha particle abundance deviates from its equilibrium value and the abundance distribution of heavy-nuclei is still under the reaction equilibrium.
We call the abundance as ``$\alpha$QSE abundance" and compare to the results of network calculations in the following section.

\subsection{Framework}
We consider a system in the equilibrium at a temperature $T$, a density $\rho$, and an electron fraction $Y_e$.
The abundance distribution in equilibrium under such a condition is determined so that the free energy of the system, $f$, is minimized under the abundance.
Thus, we solve the abundances of the nuclei such that the free energy is stationary under any infinitesimal deviations of the abundance $Y_i$ of the nuclear species $i$, or,
\begin{equation}
df = \sum_{i} \mu_i dY_i = 0,\label{eq1}
\end{equation}
where $\mu_i$ is the chemical potential of the $i$-th nucleus.
Under the non-relativistic and non-degenerate situation, we can write $\mu_i$ as
\begin{equation}
\mu_i = m_ic^2 + k_{\rm B}T \ln\left[\frac{\rho Y_i}{m_u g_i}\left(\frac{2\pi\hbar^2}{m_ik_{\rm B}T}\right)^{3/2}\right], \label{eq2}
\end{equation}
where $m_i$ and $g_i=g_i(T)$ are the mass and the partition function of the nuclear species $i$, $m_u$ is the atomic mass unit, $\hbar$ is the reduced Planck constant, and $k_{\rm B}$ is the Boltzmann constant.

In solving the ``normal" QSE abundances formulated in \cite{1998ApJ...498..808M}, we impose three constraints on the condition of the minimum free energy, i.e., the conservation of mass, the charge neutrality, the fixed abundances of heavy-nuclei $Y_h$.
In order to obtain the $\alpha$QSE abundances, we must add the fourth constraint, that the alpha particle abundance $Y_\alpha$ is fixed.
Using the fourth constraint, the above constraints can be reduced to three equations as
\begin{eqnarray}
\sum_{i\neq \alpha} Z_i Y_i &=& Y_e - 2 Y_\alpha,\label{eq3}\\
\sum_{i\neq \alpha} N_i Y_i &=& 1 - Y_e - 2 Y_\alpha ,\label{eq4}\\
\sum_{i\neq n,p,\alpha} Y_i &=& Y_h, \label{eq5}
\end{eqnarray}
where $Z_i$ and $N_i$ are proton and neutron numbers of the $i$-th nucleus.
Using Equation (\ref{eq3}), we rearrange the equation of mass-conservation to Equation (\ref{eq4}), i.e., the condition for constant neutron number in the system.
Note that $Y_\alpha$ is now one of the parameters which control the abundance distribution.
In order to solve Equation (\ref{eq1}) under the constraints (\ref{eq3})-(\ref{eq5}), we use the method of Lagrange multiplier.
As the result, the abundance of the nuclear species $i$ is written as
\begin{eqnarray}
Y_i \approx g_i A_i^{3/2} \left(\frac{\rho/m_u}{\theta}\right)^{A_i-1} \Lambda Y_n^{N_i} Y_p^{Z_i}\ e^{B_i/k_{\rm B}T} \ (i\neq n,p,\alpha),\nonumber\\\label{eq:6}
\end{eqnarray}
where $A_i$ and $B_i=(N_i m_n+Z_i m_p -m_i)c^2$ are the mass-number and the binding energy of the $i$-th nucleus, $\theta = (m_u k_{\rm B}T/2\pi \hbar^2)^{3/2}$, $\Lambda=e^{\lambda_h/k_{\rm B}T}$, and $\lambda_h$ is a Lagrange multiplier defined in Appendix \ref{app:a}.
Here we use the approximation $m_i\approx A_i m_u$.
We obtain the $\alpha$QSE abundance solving for $\Lambda$, $Y_n$, and $Y_p$ so that they satisfy Equations (\ref{eq3})-(\ref{eq5}).
Thus, they are the functions of the parameters in the system, i.e., $\rho$, $T$, $Y_e$, $Y_h$, and $Y_\alpha$.
Details for the derivation of the $\alpha$QSE abundance are given in Appendix \ref{app:a}.

\subsection{The $\alpha$QSE Abundance Distribution}
\begin{figure}[htbp]
\includegraphics[width=\hsize]{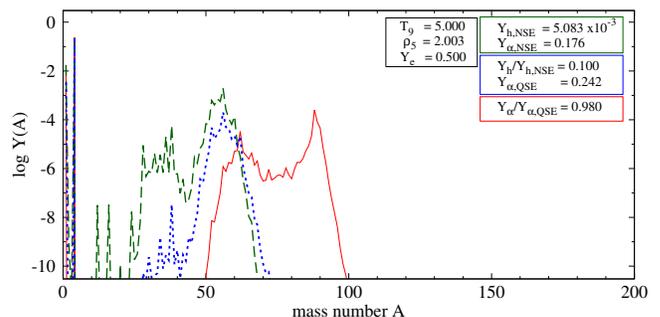}
\caption{
Abundance distributions in the NSE (green-dashed line), the normal QSE (blue-dotted line), and the $\alpha$QSE (red-solid line) in the case of $T_9=5$, $\rho_5 \approx 2$, and $Y_e=0.50$.
For the normal QSE, we set the heavy-element abundance $Y_h=0.1 Y_{h,{\rm NSE}}$.
For the $\alpha$QSE, we set the same $Y_h$ as the normal QSE, and set $Y_\alpha=0.98 Y_{\alpha,{\rm QSE}}$.
}
\label{fig:S140_T5_ye0.50_ya0.98}
\end{figure}

Here we show the $\alpha$QSE abundance distribution in some cases.
In Figure \ref{fig:S140_T5_ye0.50_ya0.98}, we show the abundance as a function of the mass-number in the nuclear statistical equilibrium (NSE), the normal QSE \citep[formulated in][]{1998ApJ...498..808M}, and the $\alpha$QSE of $T_9=T/(10^9\ {\rm K})=5$, $\rho_5=\rho/(10^5{\rm g\ cm^{-3}}) \approx2$ (corresponding to the situation of the entropy per baryon $s/k_{\rm B}=140$), and $Y_e=0.50$.
We see that the NSE abundance (the green-dashed line) peaks at iron-group nuclei.
In calculating the normal QSE abundance, we set the heavy-element abundance $Y_h=0.1 Y_{h,{\rm NSE}}$, where $Y_{h,{\rm NSE}}$ is the heavy-element abundance in the NSE of the same $T$, $\rho$, and $Y_e$.
We see that the abundance pattern does not change drastically.

Regarding the $\alpha$QSE abundance (the red-solid line), we use the values $Y_h=0.1 Y_{h,{\rm NSE}}$ and $Y_\alpha=0.98 Y_{\alpha,{\rm QSE}}$, where $Y_{\alpha,{\rm QSE}}$ is the alpha particle abundance in the normal QSE.
The abundance pattern drastically changes from the NSE and the normal QSE, and another peak appears at higher mass-number ($A\sim 90$, the $N=50$ magic nuclei).
The only 2\% lack of $Y_\alpha$ for its normal QSE value also raises the abundances of protons and neutrons to $Y_p^{\rm \alpha QSE}\approx 6.9 \times10^{-3}$ and $Y_n^{\rm \alpha QSE}\approx 1.4 \times10^{-3}$, which is 2.4 and 2.6 times larger than those of the normal QSE.
The increase of free nucleons makes the abundance distribution in the reaction equilibrium shift to higher mass-number.

\begin{figure}[htbp]
\includegraphics[width=\hsize]{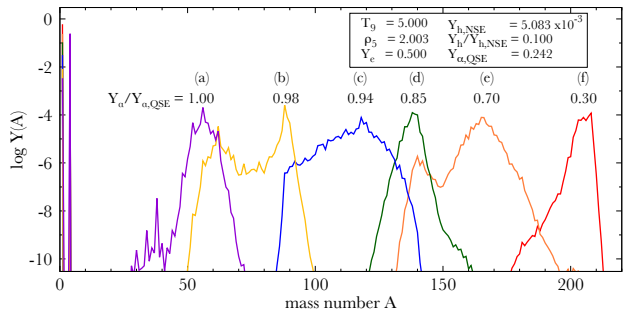}
\caption{
$\alpha$QSE abundance patterns for $Y_\alpha/Y_{\alpha,{\rm QSE}}=$1.00, 0.98, 0.94, 0.85, 0.70, 0.30.
The other parameters are the same as those in Figure \ref{fig:S140_T5_ye0.50_ya0.98}.
}
\label{fig:S140_T5_ye0.50}
\end{figure}

In Figure \ref{fig:S140_T5_ye0.50}, we show the $\alpha$QSE abundances for six cases; $Y_\alpha/Y_{\alpha,{\rm QSE}}=$ 1.00(a), 0.98(b), 0.94(c), 0.85(d), 0.70(e), and 0.30(f).
Note that the case (a) corresponds to the normal QSE.
The peaks in cases (a), (b), and (c) correspond to the iron-peak nuclei, $N=50$ magic nuclei, and $Z=50$ magic nuclei, respectively.
The peaks for cases (d), (e), and (f) indicate the locations of the $N=82$ magic number, stable nuclei due to small deformation, and $Z=82$, $N=126$ magic numbers.
We see that, with sufficiently low $Y_\alpha$, the nuclei of $A\sim 200$ can be produced in the $\alpha$QSE.

Briefly we discuss the dependence of the $\alpha$QSE abundance on the other parameters.
If we set lower $Y_h$ fixing the other parameters, the abundance peak shifts to the higher mass-number.
This is because the number of nucleons per heavy nucleus gets larger.
If we set higher entropy, the abundance peak appears at smaller-mass-number since, at the same temperature, the density and nucleon-capture reaction rates get smaller.
Normally, the larger $Y_e$ is, the smaller the peak-mass-number becomes, because adding a proton to a nucleus releases less energy than adding a neutron.
The responses of the average mass-number of the distribution to the change of parameters are given in Appendix \ref{app:b}.

\section{Conditions for $\alpha$QSE Nucleosynthesis}
In the previous section, we see that, if the alpha particle abundance $Y_\alpha$ lacks for the value in equilibrium with nucleons, large-mass-number nuclei can be produced in the $\alpha$QSE distribution.

There are two causes by which the alpha particle abundance deviates from its equilibrium value.
One is high entropy.
The high entropy decreases the abundances of light nuclei such as $^2$H, $^3$H, and $^3$He, which are well in equilibrium with free nucleons \citep[as mentioned in][]{2002PhRvL..89w1101M}.
At a high temperature ($T_9\gtrsim5$), alpha particles are produced mainly from these light nuclei, so in the high entropy environment, the reaction flow to alpha particle cannot get large enough to maintain the alpha particles in equilibrium with the nucleons.
The other is fast expansion.
At a low temperature, the production of alpha particles occurs via catalysis of heavy nuclei \citep[][]{2002PhRvL..89w1101M}, and the fast expansion (with high entropy) prevents the the production of heavy nuclei via triple-alpha reactions.

Therefore, the $\alpha$QSE abundance would appear in a high-entropy and fast-expanding material.
In this section we compare the network calculation on a simple model of expanding material to the $\alpha$QSE abundances, and discuss the condition for the $\alpha$QSE nucleosynthesis.

\subsection{The Expansion Model and Nucleosynthesis Calculation}
In order to investigate the $\alpha$QSE condition, we employ the simple model of spherically symmetric, homogeneous, and adiabatic expansion with a constant velocity.
The time-evolution of the density is
\begin{equation}
\rho(t) = \rho_0(1+t/\tau_{\rm exp})^{-3},
\end{equation}
where $\rho_0=\rho(t=0)$ is the initial mass density and $\tau_{\rm exp}=r_0/v_0$ is the expansion timescale ($r_0$ and $v_0$ is the initial radius of the material and the expansion velocity).

We use the equation of state by \cite{2000ApJS..126..501T} to obtain the entropy as the function of $\rho$, $T$, and $Y_e$, which includes the entropy of nucleons, electrons, positrons and photons.
The temperature at a time $t$ is obtained using $\rho(t)$ with the condition for the constant entropy.

Using the above expansion model, we perform the nucleosynthesis calculation.
We use the same nuclear reaction network and reaction rates in \cite{2015ApJ...810..115F} (see this paper for details).
However, we do not consider the interactions of neutrinos, e.g., $p+\bar{\nu}_e\rightarrow n+e^+$  reactions, the effects of which will be discussed in the following section.

Parameters which determine the final abundance pattern are the entropy $s$, the expansion timescale $\tau_{\rm exp}$, and the electron fraction $Y_e$ of the material at the beginning of the calculation, where $T_9=9$.

\subsection{The Dependence on the Entropy and the Expansion Timescale }
\begin{figure}[t]
   \includegraphics[width=\hsize]{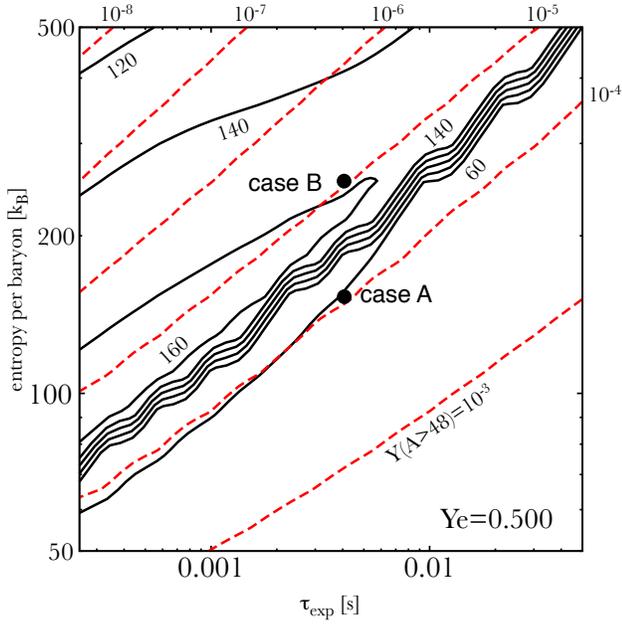}
\caption{
Contours of the average mass number of heavy-nuclei $\left<A\right>$ (black-solid lines) and the abundance of the heavy-nuclei $Y_{A>48}$ (red-dashed lines) on $s$-$\tau_{\rm exp}$ plane in the case of $Y_e=0.5$.
The contour interval is 20.
The points ``case A" and ``case B" indicate the expansions with $(s,\tau_{\rm exp})=(153,4.07\ {\rm ms})$ and $(254,4.07\ {\rm ms})$, respectively.
}
\label{fig:Ye0.50_helm}
\end{figure}

We calculate the nucleosynthesis on trajectories of each parameter set in the $s$-$\tau_{\rm exp}$ plane, in Figure \ref{fig:Ye0.50_helm}, where we set $Y_e=0.5$.
In this figure, we show the contours of the average mass number of the heavy-nuclei (black-solid line),
\begin{equation}
\left<A\right> = \frac{1}{Y_{A>48}} \sum_{A_i>48} A_i Y_i,
\end{equation}
where we define the abundance of the heavy-nuclei $Y_{A>48}$ as $Y_{A>48} = \sum_{A_i>48} Y_i$, the contours of which (red-dashed line) are also shown in the same figure.
We see that the average mass-number gets larger than 60 in the expansion where $Y_{A>48}\lesssim 10^{-5}-10^{-4}$.

\begin{figure}[htbp]
   \includegraphics[width=\hsize]{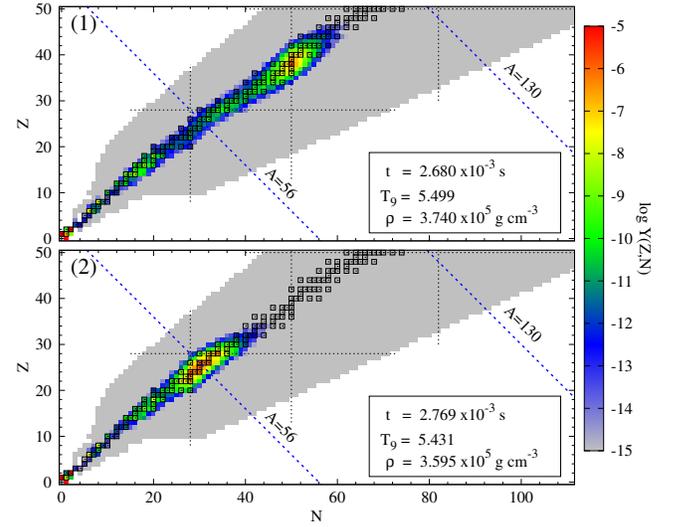}
\caption{
Snapshots of the abundance distribution in the expansion with $s=153$ and $\tau_{\rm exp}=4.07$ ms, where $T_9\approx 5.5$ (Top) and $T_9\approx 5.4$ (Bottom).
The blue-dashed lines indicate the nuclei of $A=$ 56 and 130.
Stable nuclei are shown in black squares.
}
\label{fig:cmp_00213_00053_00075}
\end{figure}

We closely investigate nucleosynthesis of some expansions in Figure \ref{fig:Ye0.50_helm}.
First, we focus on the low-entropy expansion with $s=153$ and $\tau_{\rm exp}=4.07$ ms (case A in this figure), in which $\left<A\right> <60$.
We show two snapshots of the abundance distribution (hereafter we call ``the network abundance distribution") in the nuclear chart at $T_9\approx$ 5.5 (Top) and 5.4 (Bottom) in Figure \ref{fig:cmp_00213_00053_00075}.
At $T_9\approx 6.1$, the abundance of alpha particles becomes smaller than its equilibrium value.
Then, the abundance distribution starts to shift from the normal QSE to the $\alpha$QSE one.
Therefore, the distribution concentrates around $N=50$ magic nuclei in Figure \ref{fig:cmp_00213_00053_00075}-(1).

\begin{figure}[htbp]
   \includegraphics[width=\hsize]{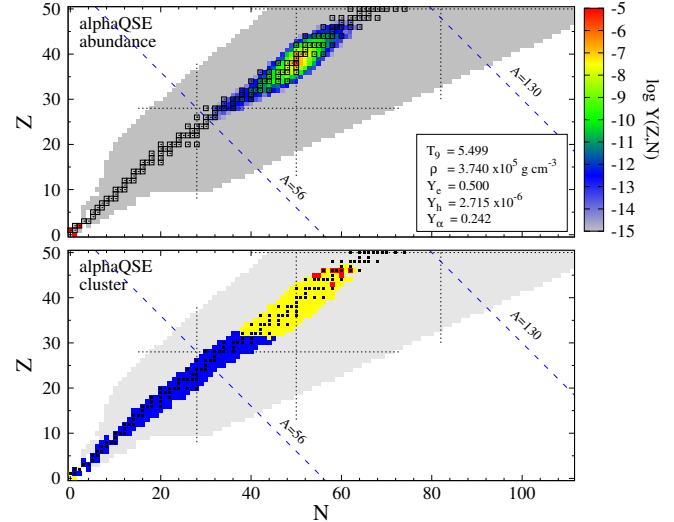}
\caption{
Top: $\alpha$QSE abundance in the same environment as Figure \ref{fig:cmp_00213_00053_00075}-(1).
Vertical and horizontal dotted lines indicate the nuclei of $Z,\ N=$28, 50, 82, and 126, i.e., those having closed proton or neutron shell.
Bottom: Extent of the $\alpha$QSE.
The nuclei of $Y(Z,A)>10^{-15}$ are shown in blue, yellow, or red.
The nuclei shown in yellow indicate those nearly in the $\alpha$QSE with the most abundant nuclear species (i.e., $0.1<r(Z,A)<10$).
Red (blue) squares indicate the nuclei whose abundance are smaller (larger) than those of $\alpha$QSE values.
}
\label{fig:qse_clstr_00213_00053}
\end{figure}

In the top panel of Figure \ref{fig:qse_clstr_00213_00053}, we show the $\alpha$QSE abundance in the same environment, i.e., at the same $T$, $\rho$, $Y_e$, $Y_h$, and $Y_\alpha$ as the top panel of Figure \ref{fig:cmp_00213_00053_00075}.
The $\alpha$QSE abundance distribution is the distribution to which the abundance tends to evolve, because the $\alpha$QSE abundance distribution is the one that minimizes the free energy under the given constraints.
We see that the network distribution, especially in large-$A$ region, is well described by the $\alpha$QSE distribution.

In order to see the extent of the $\alpha$QSE cluster in the network calculation, we define a combination
\begin{eqnarray}
r(Z,A) = \frac{Y_{\rm network}(Z,A)/Y_{\rm network}(Z_{\rm peak},A_{\rm peak})}{Y_{\rm \alpha QSE}(Z,A)/Y_{\rm \alpha QSE}(Z_{\rm peak},A_{\rm peak})},
\end{eqnarray}
where $(Z_{\rm peak},A_{\rm peak})$ indicates the the most abundant nucleus in the $\alpha$QSE abundance distribution.
If the value of $r(Z,A)$ is near the unity, it can be said that the nucleus $(Z,A)$ and $(Z_{\rm peak},A_{\rm peak})$ are in the same $\alpha$QSE cluster.
In the bottom panel of Figure \ref{fig:qse_clstr_00213_00053}, we show the extent of $\alpha$QSE cluster at $T_9\approx 5.5$, where the nuclei whose values of $r$ are between 0.1-10 are shown in yellow, indicating they are nearly in $\alpha$QSE with the nucleus $(Z_{\rm peak},A_{\rm peak})$.
From this figure, we see that the nuclei around $N=50$ magic nuclei are indeed in the $\alpha$QSE.

\begin{figure}[htbp]
   \includegraphics[width=\hsize]{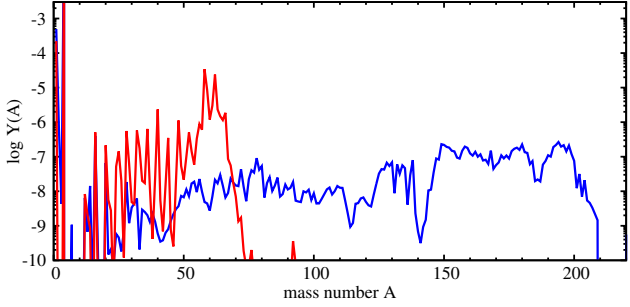}
\caption{
Final abundance distributions of the case A (red line) and case B (blue line).
}
\label{fig:lastya_00216_00213}
\end{figure}

However, at $T_9\approx 5.4$, in the network calculation, the heavy-nuclei of $A \sim 90$ are all photo-disintegrated.
This is because, in this expansion, a lot of iron-group nuclei are supplied via triple-alpha reactions.
Then the reaction cycles, ${}^{58}{\rm Fe}(n,\gamma){}^{59}{\rm Fe}(p,n){}^{59}{\rm Co}(n,\gamma){}^{60}{\rm Co}(p,n){}^{60}{\rm Ni}(n,\gamma){}^{61}{\rm Ni}$\\$(n,\alpha){}^{58}{\rm Fe}$ for example, supply alpha particles and bring $Y_\alpha$ to its equilibrium value.
Therefore, the heavy-nuclei ($A \sim 90$) in the $\alpha$QSE are photo-disintegrated and finally, the normal QSE, where iron-group nuclei are most abundant for $Y_e=0.5$, is achieved.
As the result, the final abundance peaks at iron-group nuclei as shown in Figure \ref{fig:lastya_00216_00213}.

From the above discussion, we conclude that the condition for the $\alpha$QSE nucleosynthesis, i.e., the freeze-out of the $\alpha$QSE abundance, is the production of less heavy-nuclei not to produce alpha particle sufficiently.

\begin{figure}[htbp]
   \includegraphics[width=\hsize]{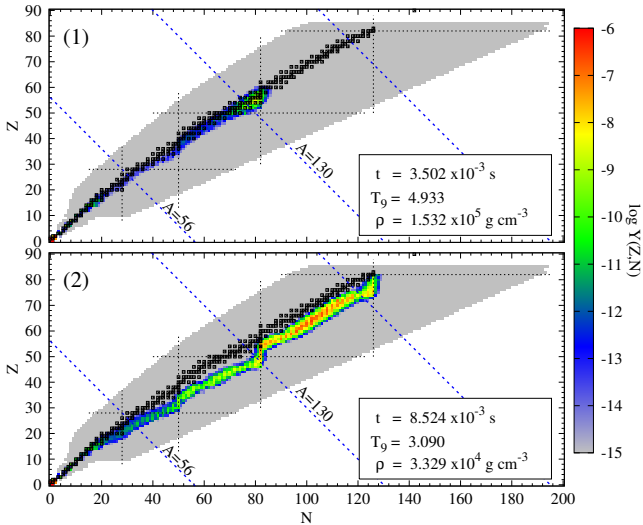}
\caption{
Snapshots of the abundance distribution in the expansion with $s=254$ and $\tau_{\rm exp}=4.07$ ms.
The temperatures of each snapshots are $T_9\approx 4.9$ and 3.1, respectively.
}
\label{fig:cmp_00216_00052_00083}
\end{figure}

\begin{figure}[htbp]
   \includegraphics[width=\hsize]{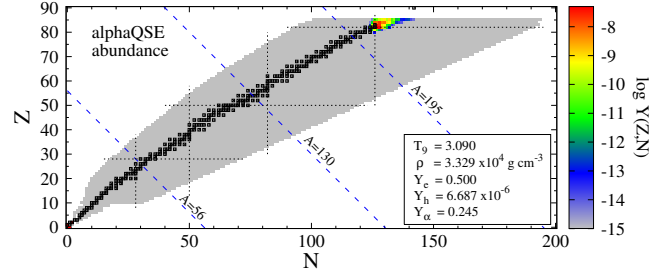}
\caption{
$\alpha$QSE abundance in the same environment as the Figure \ref{fig:cmp_00216_00052_00083}-(2).
}
\label{fig:qse_clstr_00216_00083}
\end{figure}

Next, we focus on the high-entropy expansion with $s=254$ and $\tau_{\rm exp}=4.07$ ms (``case B" in Figure \ref{fig:Ye0.50_helm}), in which $\left<A\right> =159$.
In Figure \ref{fig:cmp_00216_00052_00083}, we show two snapshots of the abundance distributions at $T_9 \approx$ 4.9 (Top) and 3.1 (Bottom) in the expansion.
In this case, the abundance of alpha particles starts to deviate from its normal QSE value at $T_9\approx 6.5$ and the abundance distribution starts to evolve to the $\alpha$QSE one.
At $T_9\approx 4.9$ (Figure \ref{fig:cmp_00216_00052_00083}-(1)), the nuclei around $N=82$ magic nuclei are indeed in the $\alpha$QSE.
Below this temperature, the abundance distribution concentrates in the neutron-rich region, as we see in Figure \ref{fig:cmp_00216_00052_00083}-(2), where the distribution of the network calculation at $T_9\approx 3.1$ is shown.
In Figure \ref{fig:qse_clstr_00216_00083}, we show the $\alpha$QSE abundance in the same environment as Figure \ref{fig:cmp_00216_00052_00083}-(2).
We see that the $\alpha$QSE abundance, to which the network abundance tends to evolve, is concentrated in very-large mass-number nuclei (around double magic nucleus of $(Z,N)=(82,126)$ and even larger).
Therefore, the neutron-rich network abundance is the transitional abundance toward the $\alpha$QSE one.
Since they cannot catch-up with the $\alpha$QSE distribution to the end, after the phase, the distribution freezes-out.
The final abundance is shown in Figure \ref{fig:lastya_00216_00213}.
We see that the elements of $A\sim 200$ are synthesized in this nucleosynthesis process.

\cite{2002PhRvL..89w1101M} first pointed out that this nucleosynthesis occurs in the fast-expanding material.
We make it clear that, although it does not proceed via the $\alpha$QSE abundance, the nucleosynthesis process can be understood as the transient phase toward the $\alpha$QSE abundance.

\begin{figure}[htbp]
   \includegraphics[width=\hsize]{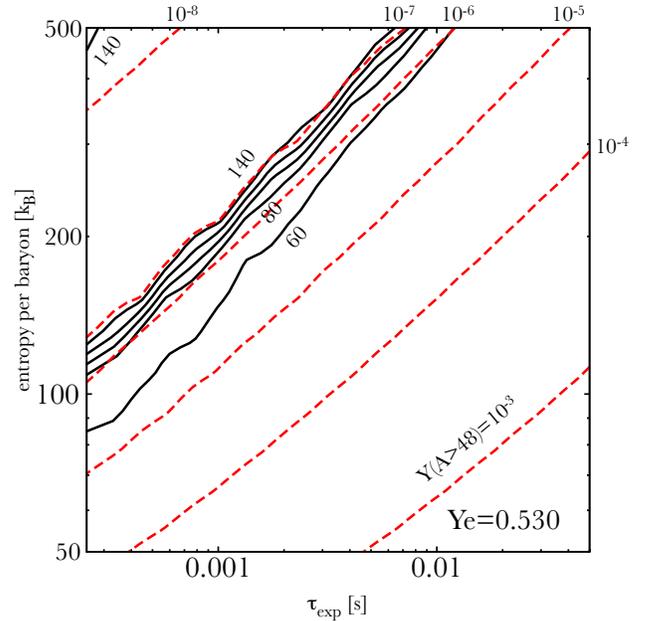}
\caption{
Same figure as Figure \ref{fig:Ye0.50_helm}, but in the case of $Y_e=0.53$.
}
\label{fig:0.53}
\end{figure}

We show the case where the initial electron fraction is $Y_e=0.53$ in the nucleosynthesis calculations in Figure \ref{fig:0.53}.
We see that the parameter region where the $\alpha$QSE nucleosynthesis occurs becomes smaller than that in the case $Y_e=0.50$.
In the case of higher electron fraction, the abundance distribution of the iron-group is concentrated in more proton-rich region.

In expansions with the same $s$ and $\tau_{\rm exp}$, increasing $Y_e$ from 0.50 to 0.53 increases the average proton number in the iron-group nuclei by $0.5-1$ at the same temperature.
Then the reaction cycle which mainly supply alpha particles changes to more proton-rich side, e.g., ${}^{56}{\rm Fe}(p,\gamma){}^{57}{\rm Co}(p,\gamma){}^{58}{\rm Ni}(n,\gamma){}^{59}{\rm Ni}(n,\alpha){}^{56}{\rm Fe}$.
In general, the cross sections of $(n,\alpha)$ reactions, which supply alpha particles, are large for proton-rich nuclei.
On the other hand, in this high temperature ($T_9\sim 5-6$), the increasing $Y_e$ at the beginning of the calculation changes the neutron abundance only by a factor.
Therefore, increasing $Y_e$ enhances the alpha particle production via $(n,\alpha)$ reactions of proton-rich iron-group nuclei.
Thus, satisfying the condition for the $\alpha$QSE nucleosynthesis becomes more difficult in the proton-rich material.

\section{$\alpha$QSE Nucleosynthesis in HN Wind}
\subsection{Comparison of the Network Abundance in HN Wind to the $\alpha$QSE Abundance}
Here we compare the results of the nucleosynthesis calculation of the HN wind with the $\alpha$QSE abundance.
We construct the spherically symmetric, steady solution of the NDW using parameter set in the first line of Table \ref{tab:modelpar}.
They are parameters of the HN wind used in \cite{2015ApJ...810..115F}.
Then, we perform the nucleosynthesis calculation on the constructed wind solution.
Now we include the (anti)neutrino-absorption reactions of nucleons.
The methods of construction of the wind and the nucleosynthesis calculation are summarized in \cite{2015ApJ...810..115F}.

\begin{table*}[htbp]
\caption{The Parameters of the HN Wind and the Ordinary SN Wind}
\begin{center}
\begin{tabular*}{\hsize}{@{\extracolsep{\fill}}lccccccccc}
\hline \hline
Model & $M$&$R_{\nu}$&$\rho_{0}$&$L_{\nu_e} $&$L_{\bar{\nu}_e}$&$L_{\nu_x}$&$\epsilon_{\nu_e}$&$\epsilon_{\bar{\nu}_e}$&$\epsilon_{\nu_x}$ \\
&$(M_\odot)$ & $({\rm km})$ & $({\rm g\ cm^{-3}})$ & \multicolumn{3}{c}{$({\rm 10^{51}erg\ s^{-1}})$} & \multicolumn{3}{c}{$({\rm MeV})$} \\ \hline
HN                 & 3.0 & 15 &$10^{11}$ & 150 & 165 & 15 & 9.00 & 11.3 & 25.0\\
Ordinary SN & 1.4 & 10 & $10^{10}$ & 1     & 1      & 1   & 12.0 & 16.0 & 14.0\\
\hline
\end{tabular*}
\end{center}
\label{tab:modelpar}
\end{table*}

\begin{figure}[htbp]
\includegraphics[width=\hsize]{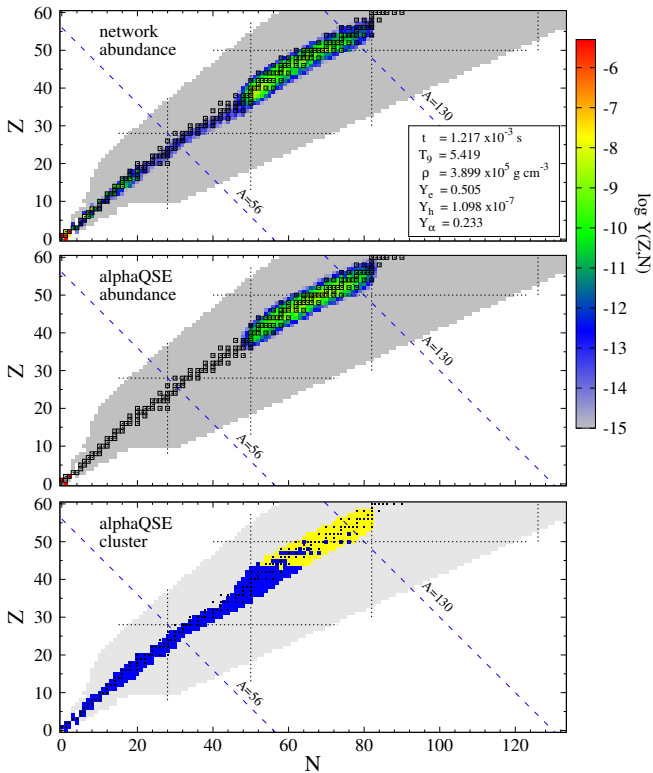}
\caption{
Top: Snapshot of the abundance distribution at $T_9\approx 5.4$ in our network calculation.
Middle: $\alpha$QSE abundance distribution at the same temperature, density, electron fraction, number of heavy-nuclei $Y_h$, and number of alpha particle $Y_\alpha$.
Bottom: Extent of the $\alpha$QSE.
The color representation is the same as the bottom panel of Figure \ref{fig:qse_clstr_00213_00053}.
}
\label{fig:qse_clstr_00007_00054}
\end{figure}

We compare the network abundance in the HN wind and the $\alpha$QSE abundance in the same environment in Figure \ref{fig:qse_clstr_00007_00054}.
From the snapshot of the abundance distribution at $T_9\approx 5.4$ in our network calculation (top panel), we see that heavy-nuclei are concentrated in the region near the stable line and $N=50-82$ and $Z=35-55$.
This trend is well described by the $\alpha$QSE abundance distribution with the same parameter set of $Y_e$, $Y_h$, and $Y_\alpha$ (middle panel).
In the bottom panel, we show the extent of the $\alpha$QSE cluster.
We see that most nuclei in the region of $N=50-82$ and $Z=35-55$ are indeed in $\alpha$QSE cluster.

\begin{figure}[t]
\includegraphics[width=\hsize]{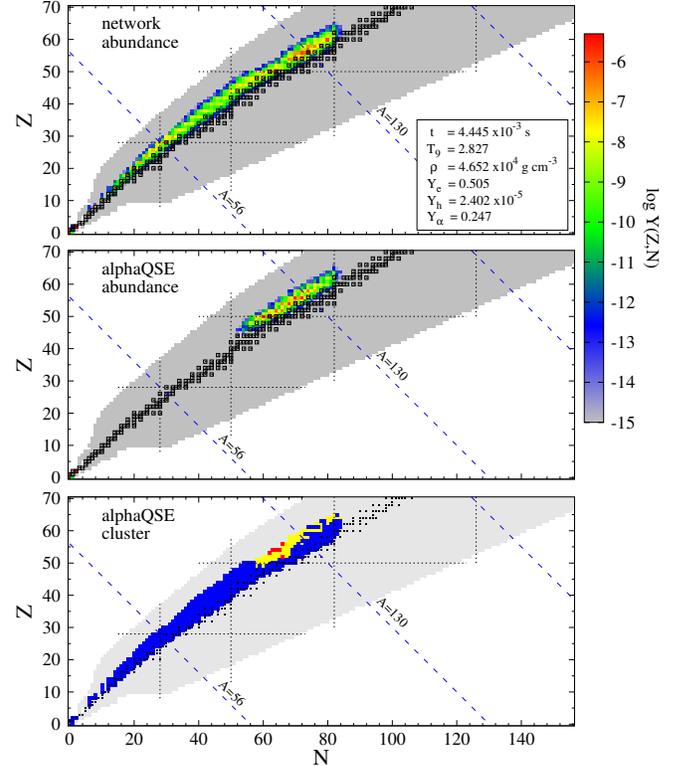}
\caption{
Same figure as Figure \ref{fig:qse_clstr_00007_00054}, but at $T_9\approx 2.8$.
}
\label{fig:qse_clstr_00007_00120}
\end{figure}

We show the snapshot at $T_9\approx2.8$ in Figure \ref{fig:qse_clstr_00007_00120}.
At this temperature, the network result shows proton-rich distribution in $N=50-82$ region (see top panel).
The $\alpha$QSE abundance distribution (middle panel) shows the same trend.
Therefore, the trends of the network abundance can be expressed by the $\alpha$QSE abundance distribution in the same environment.
After this, the reaction rates of heavy-nuclei decrease and the nuclear abundance freezes-out.

\begin{figure}[t]
\includegraphics[width=\hsize]{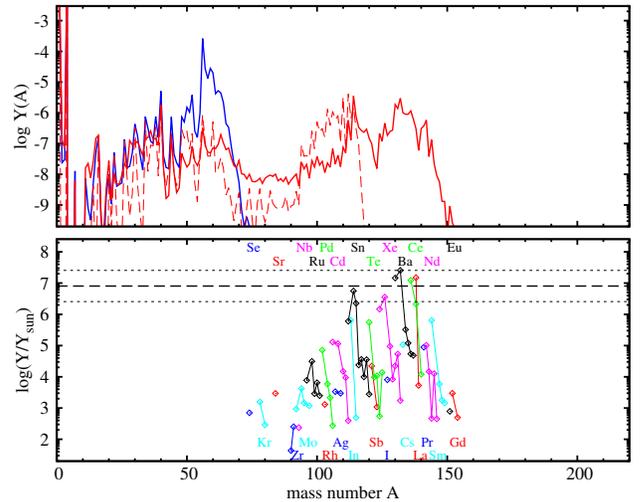}
\caption{
Top: Final abundance distributions of the HN wind (red-solid line) and the ordinary SN wind (blue-dashed line).
The thin-dashed line indicates the result of the HN wind without neutrino-interactions.
Bottom: Production factors of the HN wind.
}
\label{fig:prodfctr_ya_00007}
\end{figure}

The top panel of Figure \ref{fig:prodfctr_ya_00007} shows the final abundance distribution of the HN wind.
We see that the nuclei of $A\sim 110-140$ are mainly produced by $\alpha$QSE nucleosynthesis.
The bottom panel shows the production factors of the final products.
The $p$-nuclei of $A\sim 110-140$ are main products.

\subsection{The Effects of Neutrinos}
Here we discuss the effects of neutrinos on the $\alpha$QSE nucleosynthesis.
In the proton-rich condition, anti-electron neutrinos supply neutrons via $p+\bar{\nu}_e\rightarrow n+e^+$ reactions.
Neutrons are easily absorbed by heavy-nuclei.
Therefore, the $\alpha$QSE abundance distribution is maintained for a longer time because photo-disintegrations are prevented by neutron-absorption reactions.

In the top panel of Figure \ref{fig:prodfctr_ya_00007}, we also show the result for the nucleosynthesis calculation without neutrino interactions on the same background as the HN wind model (thin-dashed line).
We see that the nuclei of $A>120$ are photo-disintegrated to the nuclei of $A=100-120$.
In the absence of neutrons, these heavy-nuclei cannot exist against photo-disintegrations.
Therefore, the neutrino-interactions indeed maintain the $\alpha$QSE abundance.

\subsection{Differences between the NDW in SNe and HNe}
Here we discuss the differences of the nucleosynthesis between the ordinary SN wind and the HN wind.
We construct the solution of the NDW in the ordinary SN in the same way as the HN, but we use the parameter set in the second line of Table \ref{tab:modelpar}.

In the top panel of Figure \ref{fig:prodfctr_ya_00007}, we show the final abundance distribution of the ordinary SN wind (blue line).
From this figure, we see that only nuclei up to iron-group are produced in the ordinary SN wind.

\begin{figure}[htbp]
\begin{minipage}{\hsize}
\includegraphics[width=\hsize]{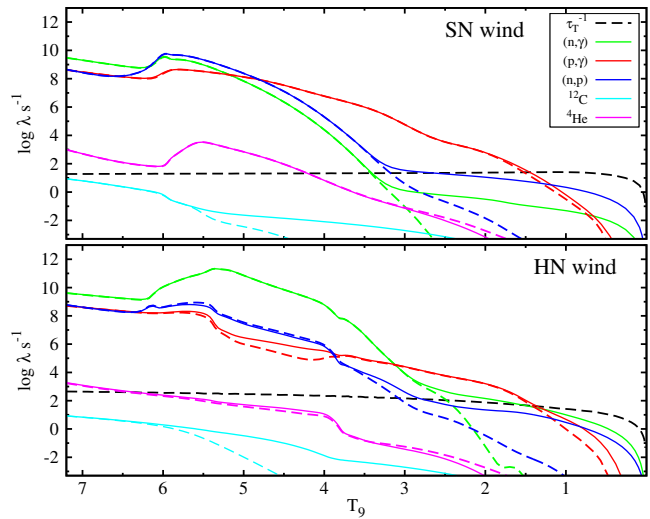}
\end{minipage}
\caption{
Top: Time evolution of the average reaction rates per nucleus in the heavy-nuclei of  $A\ge12$ in the trajectory of the ordinary SN wind.
Bottom: Same figure as the top panel, but in the case of the HN wind.
The rates of $(n,\gamma)$, $(p,\gamma)$, $(n,p)$ reactions are shown in green, red, and blue-solid lines, respectively, and their inverse reaction rates are shown in dashed lines.
The temperature-decreasing rate $\tau_T^{-1}$ is shown in the black-dashed line.
}
\label{fig:f1}
\end{figure}

The difference between final abundance distributions in the ordinary SN and the HN winds is caused by the differences in the relation between the expansion rate and the reaction rates in the winds.
We show the average rates of individual reactions per nucleus in the nuclei of $A\ge 12$ in the ordinary SN wind in the top panel of Figure \ref{fig:f1} \citep[for the definition of these rates, see Section 4 in][]{2015ApJ...810..115F}.
In the same figure, we also show the rates of reactions which produce alpha particles (pink-solid line), and reactions which produce $^{12}{\rm C}$ from alpha particles, i.e., the triple-alpha and ${}^{4}{\rm He}(\alpha n,\gamma){}^{9}{\rm Be}(\alpha,n){}^{12}{\rm C}$ reactions (light-blue-solid line).
Note that the rates of their inverse reactions are shown in dashed lines.

We see that, throughout the calculation, the rates of $^{12}$C-production and decomposition reactions are below the temperature-decreasing rate $\tau_T^{-1} = T^{-1}dT/d\tau$.
In this situation, alpha particles are not consumed and remain after the temperature decreases.
This is called ``alpha-rich freeze-out."
In addition, until the temperature deceases to $4.5\times 10^9\ {\rm K}$, the production and decomposition rates of alpha particles are larger than the temperature-decreasing rate and the both rates are almost the same.
This indicates that the alpha particles are in equilibrium with nucleons until $4.5\times 10^9\ {\rm K}$.
Note that the rates of $(n,\gamma)$, $(p,\gamma)$, and $(n,p)$ and their inverse reactions are almost the same and much larger than the temperature-decreasing rate.
Therefore, the normal QSE abundance, formulated in \cite{1998ApJ...498..808M}, is in good agreement with the network abundance.
In this case, since $Y_e$ is near 0.5, the normal QSE abundance peaks around iron-group nuclei.
As a result, after the abundance freezes-out, the iron-peak remains as shown in the top panel of Figure \ref{fig:prodfctr_ya_00007}.

In the case of the HN wind (Bottom panel of Figure \ref{fig:f1}), however, the reaction rates of alpha particles get smaller than the temperature-decreasing rate at $\sim 6\times 10^9\ {\rm K}$.
Below this temperature, the abundance of alpha particles cannot catch-up with and deviates from its equilibrium value.
Therefore, the $\alpha$QSE nucleosynthesis indeed occurs in the HN wind.

\section{discussion}
\subsection{Production of $p$-nuclei}
As shown in Section 3, the $\alpha$QSE abundance is concentrated in proton-rich region at low temperature.
As a result, as seen in the bottom panel of Figure \ref{fig:prodfctr_ya_00007}, $p$-nuclei of $A=110-140$ are mainly produced.
Heavy $p$-nuclei of $A=110-125$ are underproduced in the $\gamma$-process so, if the electron fraction of the wind is near 0.5, this HN wind model may be a possible site for the production of these $p$-nuclei.

Such an investigation with smaller parameter range was performed in \cite{2004ApJ...617L.131J}.
They pointed out that free neutrons are important for this $\alpha$QSE nucleosynthesis because they keep the abundance distribution in a high-mass one.
As mentioned in the previous section, because of the large neutrino luminosities, a lot of neutrons are supplied by $p+\bar{\nu}_e\rightarrow n+e^+$ reactions in the HN wind.
Thus the $p$-nuclei production through the $\alpha$QSE nucleosynthesis would indeed occur in the HN wind.

\subsection{Application to Other Astrophysical Phenomena}
In the previous sections, we show that the $\alpha$QSE nucleosynthesis occurs in the sufficiently high-entropy and the sufficiently fast-expanding material.
Therefore, the $\alpha$QSE nucleosynthesis would occur not only in HN winds but also some other phenomena which eject the material in such environment.

A high entropy would be achieved in the material in the jet in gamma-ray-bursts (GRBs) but, when we consider a fireball, the entropy per baryon required for the terminal Lorentz factor to be $\sim 100$ is $\sim 10^5 k_{\rm B}$.
It is too large to make heavy-elements during the expansion, like big-bang nucleosynthesis \citep{2003ApJ...588..931B}.

However, as mentioned in \cite{2003ApJ...595..294I}, in connection with GRBs, moderately baryon-rich material would be ejected as ``circum-jet outflow" around the collimated jet.
There would be the outflows of ``failed-GRB" events, when the activity of the central engine is weak or the baryon-loading process occurs more efficiently.
If the outflows are proton-rich, as we show in this paper, the $\alpha$QSE nucleosynthesis would occur in the material.

In the center of super-massive star (SMS), the mass of which is $\sim 10^5M_\odot$, the entropy would be 100-200.
One example is a 55,500 $M_\odot$ star in \cite{2014ApJ...790..162C}.
Although this star explodes as a thermonuclear SN by the explosive He burning, the central entropy becomes $\sim 200$ at the maximum central temperature.
Therefore, if the explosion occurs when the SMS collapses and the temperature of the ejecta rapidly decreases from $\sim 10^{10}$ K, the $\alpha$QSE nucleosynthesis would occur in the material.

\subsection{Electron Fraction in the Material}
The $\alpha$QSE nucleosynthesis, proposed in this paper, occurs in slightly proton-rich ($Y_e\gtrsim 0.5$) material.
In the material of $Y_e<0.5$, with such a high entropy and a high expansion velocity, the $r$-process will occur and the $\alpha$QSE abundance distribution will not achieved.
If the elemental abundance pattern of $\alpha$QSE is discovered in some metal-poor stars, it would constrain the electron fraction of the ejecta of HNe and GRBs, which reflects the circumstance and mechanism of them.

\section{conclusions}
In this work, we investigate the nucleosynthesis in a high-entropy and fast-expanding material where the alpha particle abundance becomes below the value in equilibrium with nucleons.
We construct the equilibrium abundance distribution in such an expansion, ``$\alpha$QSE," which minimizes the free energy under four constraints: the charge neutrality, the numbers of baryons, heavy nuclei, and alpha particles.
We show that the nucleosynthesis of the NDWs in HNe can be understood as the freeze-out of the $\alpha$QSE abundances.
We also investigate the dependence of the entropy $s$ and the expansion timescale $\tau_{\rm exp}$ on the $\alpha$QSE nucleosynthesis, using a simple expansion model.
We show that the $\alpha$QSE nucleosynthesis would occur not only in the NDW in HNe but also in the material with such high entropy and high expansion velocity.
We conclude that the condition for the $\alpha$QSE nucleosynthesis is that the less production of heavy-nuclei which are catalyses for the supply of alpha particles.

\acknowledgements
We thank Takashi Nakamura, Takahiro Tanaka, and Masaru Shibata for fruitful discussions.
S.F. is supported by Research Fellowship of Japan Society for the Promotion of Science (JSPS) for Young Scientists (No.26-1329). 
This work has been supported by Grant-in-Aid for Scientific Research (No.23740160, No.24244028, and No.25103512) and by the HPCI Strategic Program of Japanese MEXT.

\bibliographystyle{apj}
%\expandafter\ifx\csname natexlab\endcsname\relax\def\natexlab#1{#1}\fi
\bibliography{apj-jour,reference}

\appendix

\section{Derivation of the $\alpha$QSE Abundance} \label{app:a}
Here we explain details of the formulation of the $\alpha$QSE abundance.
The $\alpha$QSE abundance minimizes the free energy (i.e., Equation \ref{eq1}) under the constraints of Equations (\ref{eq3})-(\ref{eq5}).
We obtain such an abundance using the method of Lagrange multiplier.
We find the abundance for that the function
\begin{equation}
f + \lambda_n\left(1-Y_e - \sum_{i\neq \alpha}N_iY_i\right) +\lambda_p\left(Y_e - \sum_{i\neq \alpha}Z_iY_i\right) + \lambda_h \left(Y_h - \sum_{i\neq n,p,\alpha}Y_i \right) \label{eq:a1}
\end{equation}
is stationary and that satisfies Equations (\ref{eq3})-(\ref{eq5}).
Then we find the condition that the function (\ref{eq:a1}) is stationary under any infinitesimal change of the abundance as
\begin{eqnarray}
(\mu_n-\lambda_n)dY_n+(\mu_p-\lambda_p)dY_p + \sum_{i\neq n,p,\alpha}(\mu_i - N_i\lambda_n - Z_i \lambda_p-\lambda_h)dY_i = 0 \label{eq:a2}
\end{eqnarray}
Note that we now consider the abundance of alpha particles $Y_\alpha$ as one of the parameter in the system.
Therefore, we do not consider the infinitesimal change of alpha particles (i.e., $dY_\alpha=0$) when we take the derivative of (\ref{eq:a1}).
From Equation (\ref{eq:a2}), we obtain the relations
\begin{eqnarray}
\mu_i =
\begin{cases}
\lambda_i \hspace{10mm} & (i=n,p),\\
N_i \mu_n + Z_i \mu_p + \lambda_h \hspace{10mm} & (i\neq n,p,\alpha). \label{eq:a3}
\end{cases}
\end{eqnarray}
Solving for $Y_i$ from Equations (\ref{eq2}) and (\ref{eq:a3}), we finally obtain Equation (\ref{eq:6}).

\section{the response of the average mass-number of $\alpha$QSE abundance to the change of parameters}\label{app:b}
Here we discuss the dependence of the $\alpha$QSE abundance to the parameters in the system.
The following formula are convenient for investigating the dependence of the average mass-number of heavy-elements on the parameters.
First, we consider the response to the change of $Y_\alpha$.
Taking the derivative of the constraints (\ref{eq3})-(\ref{eq5}) with respect to $Y_\alpha$, we obtain 
\begin{eqnarray}
-2 &=& Y_p \frac{\partial \ln Y_p}{\partial Y_\alpha} + \sum_{i\neq n,p,\alpha} Z_i \frac{\partial Y_i}{\partial Y_\alpha}, \label{eq:b1}\\
-2 &=& Y_n \frac{\partial \ln Y_n}{\partial Y_\alpha} + \sum_{i\neq n,p,\alpha} N_i \frac{\partial Y_i}{\partial Y_\alpha}, \label{eq:b2}\\
0  &=& \sum_{i\neq n,p,\alpha} \frac{\partial Y_i}{\partial Y_\alpha}. \label{eq:b3}
\end{eqnarray}
From the relation (\ref{eq:6}), we find
\begin{equation}
\frac{\partial Y_i}{\partial Y_\alpha} = Y_i \left(Z_i \frac{\partial \ln Y_p}{\partial Y_\alpha} + N_i \frac{\partial \ln Y_n}{\partial Y_\alpha}
 + \frac{\partial \ln \Lambda}{\partial Y_\alpha} \right),
\end{equation}
and using this, Equations (\ref{eq:b1})-(\ref{eq:b3}) become
\begin{eqnarray}
-2 &=& \left(Y_p +\left<Z^2\right>_h Y_h \right) \frac{\partial \ln Y_p}{\partial Y_\alpha} + \left<ZN\right>_h Y_h                                 \frac{\partial \ln Y_n}{\partial Y_\alpha} + \left<Z\right>_h Y_h \frac{\partial \ln \Lambda}{\partial Y_\alpha}, \label{eq:b5}\\
-2 &=& \left<ZN\right> Y_h                                     \frac{\partial \ln Y_p}{\partial Y_\alpha} + \left(Y_n + \left<N^2\right>_h Y_h \right)\frac{\partial \ln Y_n}{\partial Y_\alpha} + \left<N\right>_h Y_h \frac{\partial \ln \Lambda}{\partial Y_\alpha}, \label{eq:b6}\\
0  &=& \left<Z\right>_h                                            \frac{\partial \ln Y_p}{\partial Y_\alpha} + \left<N\right>_h                                           \frac{\partial \ln Y_n}{\partial Y_\alpha} +                                      \frac{\partial \ln \Lambda}{\partial Y_\alpha}, \label{eq:b7}
\end{eqnarray}
where $\left<\cdot \right>_h$ denotes the average taken using the nuclei except $n,p,\alpha$, for example,
\begin{eqnarray}
\left<Z\right>_h = \frac{1}{Y_h}\sum_{i\neq n,p,\alpha} Z_i Y_i.
\end{eqnarray}
The derivative of $\left<A\right>_h$ is
\begin{eqnarray}
\frac{\partial \left<A\right>_h}{\partial Y_\alpha} = \left<ZA\right>_h \frac{\partial \ln Y_p}{\partial Y_\alpha} + \left<NA\right>_h \frac{\partial \ln Y_n}{\partial Y_\alpha} + \left<A\right>_h \frac{\partial \ln \Lambda}{\partial Y_\alpha}. \label{eq:b9}
\end{eqnarray}

Solving Equations (\ref{eq:b5})-(\ref{eq:b7}) for $\partial \ln Y_p/\partial Y_\alpha$, $\partial \ln Y_n/\partial Y_\alpha$, and $\partial \ln \Lambda/\partial Y_\alpha$, and substituting these into Equation (\ref{eq:b9}), we obtain the response of $\left<A\right>_h$ to the change of $Y_\alpha$ as
\begin{equation}
\frac{\partial \left<A\right>_h}{\partial Y_\alpha} = - \frac{4}{DY_h}\left[\left(\sigma_{ZZ}^2 \sigma_{NN}^2 - \sigma_{ZN}^4\right) + \left(\sigma_{ZZ}^2+\sigma_{ZN}^2\right)\frac{Y_n}{2Y_h} + \left(\sigma_{NN}^2+\sigma_{ZN}^2\right)\frac{Y_p}{2Y_h}\right], \label{eq:b10}
\end{equation}
where we define the variances of $Z$ and $N$ of heavy nuclei, and the quantity $D$ as
\begin{eqnarray}
\sigma_{ZZ}^2 &=& \left< Z^2 \right>_h - \left< Z\right>_h^2,\\
\sigma_{NN}^2 &=& \left< N^2 \right>_h - \left< N\right>_h^2,\\
\sigma_{ZN}^2 &=& \left< ZN \right>_h - \left< Z\right>_h\left< N\right>_h,\\
D &=& \left(\sigma_{ZZ}^2 + \frac{Y_p}{Y_h}\right)\left(\sigma_{NN}^2 + \frac{Y_n}{Y_h}\right) - \sigma_{ZN}^4. \label{eq:b14}
\end{eqnarray}
Note that the variance $\sigma_{ZZ}^2$ ($\sigma_{NN}^2$) is large when the abundance distribution spreads in $Z$ ($N$) space.
On the other hand, the covariance $\sigma_{ZN}^2$ is large when the distribution tightly correlates in the $N$-$Z$ plane.
In our case, $\sigma_{ZN}^2$ is positive because the abundance distribution generally shows the positive correlation on the chart of nuclides.
These variants satisfy the relation $\left| \sigma_{ZN}^2\right| \le \sigma_{ZZ}\sigma_{NN}$ so that we have $D>0$ and thus, $\partial \left<A\right>_h/\partial Y_\alpha<0$.

In the case where the abundance distribution spreads enough so that $\sigma^2 \gg Y_{\rm nuc}/Y_h$, where $Y_{\rm nuc}$ is the abundance of the nucleons, the first term in the square bracket in Equation (\ref{eq:b10}) dominates other terms and $D\approx \sigma_{ZZ}^2 \sigma_{NN}^2 - \sigma_{ZN}^4$.
Therefore, in this case, $\partial \left<A\right>_h/\partial Y_\alpha \approx -4/Y_h$, which indicates that the nucleons supplied by the decrease of $Y_\alpha$ are all consumed by heavy-nuclei.
In this case, the small change $dY_\alpha \sim Y_h$ is enough to increase $\left<A\right>_h$.

On the other hand, when the nuclear distribution concentrates near some magic nuclei, the values of $\sigma^2 \sim 1$ are usually smaller than $Y_{\rm nuc}/Y_h$, the last two terms in the square bracket dominate.
At the same time, $D\approx Y_pY_n/Y_h^2$ and we find $\partial \left<A\right>_h/\partial Y_\alpha \sim -\sigma^2/Y_{\rm nuc}$, which indicates the supplied nucleons do not used to increase $\left<A\right>_h$ and remain free.
In this case, $\left<A\right>_h$ gets large only for the change of the abundance of alpha particles $dY_\alpha \gtrsim Y_{\rm nuc}\gg Y_h$.

In the same way, we derive the dependence of $\left<A\right>_h$ on the other parameters.
The responses to the changes of $\rho$, $Y_e$, $Y_h$ are the follows:
\begin{eqnarray}
\frac{\partial \left<A\right>_h}{\partial \ln \rho} &=& \frac{1}{DY_h}\left[\left(\sigma_{ZZ}^2 \sigma_{NN}^2 - \sigma_{ZN}^4\right)(Y_p+Y_n)+\sigma_{AA}^2 \frac{Y_pY_n}{Y_h}\right],\label{eq:b15}\\
\frac{\partial \left<A\right>_h}{\partial Y_e} &=& \frac{1}{DY_h}\left[\left(\sigma_{ZZ}^2+\sigma_{ZN}^2\right)\frac{Y_n}{Y_h}-\left(\sigma_{NN}^2+\sigma_{ZN}^2\right)\frac{Y_p}{Y_h}\right],\label{eq:b16}\\
\frac{\partial \left<A\right>_h}{\partial Y_h} &=& \frac{1}{DY_h}\left[-\left(\sigma_{ZZ}^2 \sigma_{NN}^2 - \sigma_{ZN}^4\right)\left<A\right>_h+\left(\sigma_{ZZ}^2-\sigma_{ZN}^2\right)\frac{Y_n}{Y_h}\left<Z\right>_h+\left(\sigma_{NN}^2-\sigma_{ZN}^2\right)\frac{Y_p}{Y_h}\left<N\right>_h\right],\label{eq:b17}
\end{eqnarray}
where $D$ is defined in Equation (\ref{eq:b14}).

\end{document}